\documentclass[journal,comsoc]{IEEEtran}
\usepackage[T1]{fontenc}

\usepackage{cite}

\ifCLASSINFOpdf
  \usepackage[pdftex]{graphicx}
\else
\fi
\usepackage{amsmath}
\usepackage[cmintegrals]{newtxmath}
\usepackage{algorithm}
\floatname{algorithm}{Pseudo-code}
\usepackage[noend]{algpseudocode}

\usepackage{cite}
\usepackage{textcomp}
\usepackage{url}
\usepackage{subcaption}
\usepackage{autobreak}
\usepackage{tabularx}
\usepackage{multirow}
\usepackage{listings}

\lstdefinestyle{bash}{
  language=bash,
  basicstyle=\ttfamily\small,
  keywordstyle=\color{blue!70},
  commentstyle=\color{gray},
  stringstyle=\color{olive},
  showstringspaces=false,
  breaklines=true,
  postbreak=\mbox{\textcolor{red}{$\hookrightarrow$}\space},
  numbers=left,
  numberstyle=\color{gray},
  stepnumber=1,
  firstnumber=1,
  numbersep=5pt,
}

\newcommand{\V}{\mathcal{V}}
\newcommand{\E}{\mathcal{E}}

\newcommand{\bm}[1]{\text{\boldmath $#1$}}

\algdef{SE}[DOWHILE]{Do}{doWhile}{\algorithmicdo}[1]{\algorithmicwhile\ #1}%

\usepackage[usenames,dvipsnames,svgnames,table]{xcolor}
\usepackage{comment}

\newcommand{\Fin}{\color{black}}

\newif\iftoc
\tocfalse
\usepackage{CJKutf8}

\def\BibTeX{{\rm B\kern-.05em{\sc i\kern-.025em b}\kern-.08em
    T\kern-.1667em\lower.7ex\hbox{E}\kern-.125emX}}

\begin{document}
%
\title{Service Function Chaining Architecture for \\Multi-hop Split Inference and Learning}
%
%
%

\author{Takanori~Hara
        and~Masahiro~Sasabe
\thanks{T.~Hara is with the Division of Information Science, Nara Institute of Science and Technology, Nara, Japan, e-mail: hara@ieee.org. M.~Sasabe is with the Faculty of Informatics, Kansai University, Takatsuki, Japan, email: m-sasabe@ieee.org.}}

\maketitle

\begin{abstract}
  Service Function Chaining (SFC) is a networking technique that ensures traffic traverses a predefined sequence of service functions, realizing arbitrary network services through dynamic and efficient communication paths.
  Inspired by this concept, we propose an SFC-based architecture for Multi-hop Split Inference (MSI), where split sub-models are interpreted as service functions and their composition forms a service chain representing the global model.
  By leveraging SFC, the proposed architecture dynamically establishes communication paths for split sub-models, ensuring efficient and adaptive execution.
  Furthermore, we extend this architecture to Multi-hop Split Learning (MSL) by applying SFC to the bidirectional communication required for training tasks.
  To realize the proposed architecture, we design Neural Service Functions (NSFs) to execute split sub-models as transparent TCP proxies and integrate them with Segment Routing over IPv6 (SRv6) and the extended Berkeley Packet Filter (eBPF)-based SFC proxy. This integration ensures efficient ML processing over dynamic routing while maintaining compatibility with existing applications.
  Evaluation results demonstrate that (1) the proposed architecture is feasible for both MSI and MSL; (2) it is particularly suitable for real-time inference in MSI scenarios with small mini-batch sizes; (3) it supports dynamic path reconfiguration, enabling adaptive responses to changing network conditions while minimizing the impact of control mechanisms on inference and learning processes.
\end{abstract}
\begin{IEEEkeywords}
  Service Function Chaining (SFC), Multi-hop Split Inference/Learning (MSI/MSL), Neural Service Function (NSF), Segment Routing over IPv6 (SRv6), transparent proxy (TPROXY), extended Berkeley Packet Filter (eBPF).
\end{IEEEkeywords}
\Fin
\section{Introduction}
\label{sec:Introduction}
Service Function Chaining (SFC)~\cite{rfc7665} is a networking technique that ensures traffic traverses a predefined sequence of service functions, enabling the realization of diverse network services through dynamic and efficient communication paths.
This method facilitates the establishment of end-to-end logical communication paths (i.e., service paths) and their aggregation into network slices, which are isolated and tailored to meet specific service requirements.
These functionalities are supported by Network Functions Virtualization (NFV) and Software Defined Networking (SDN).
Network slicing use cases are generally classified into three main categories: enhanced Mobile BroadBand (eMBB), massive Machine Type Communication (mMTC), and Ultra-Reliable Low Latency Communication (URLLC).

In the context of emerging 6G systems, the integration of AI/ML-based inference and training workloads is expected to generate massive data volumes and impose significant demands on network resources.
These challenges, coupled with stringent data privacy regulations such as GDPR~\cite{chassang17ImpactEUGeneral}, necessitate the adoption of distributed machine learning (ML) approaches that eliminate the need for sharing raw data~\cite{joshiSplitfedLearningClientSide2021,linHierarchicalSplitFederated2024}.
Recent advancements in distributed ML paradigms, including SplitFed Learning (SFL)~\cite{joshiSplitfedLearningClientSide2021} and Multi-hop Parallel Split Learning (MP-SL)~\cite{tiranaMPSLMultihopParallel2024}, have demonstrated improvements in both model privacy and resource efficiency.
However, these paradigms typically rely on static communication paths, which limit their ability to adapt to dynamic network conditions.

To address these challenges, this paper proposes a novel architecture that integrates SFC~\cite{rfc7665} with Multi-hop Split Inference (MSI) and Multi-hop Split Learning (MSL).
This architecture interprets split sub-models as service functions, with their composition forming a service chain that represents the global model.
By leveraging the inherent similarity between SFC and MSI, the proposed architecture dynamically establishes service paths for split sub-models, ensuring efficient and secure communication while adapting to changing network conditions.
Furthermore, since MSL requires bidirectional communication for training tasks, the architecture extends the application of SFC to manage return traffic effectively.

The proposed architecture leverages three key technologies: Segment Routing over IPv6 (SRv6)~\cite{rfc8986}, transparent proxy (TPROXY)~\cite{tproxy}, and extended Berkeley Packet Filter (eBPF)-based SFC proxy~\cite{haeberleCachingSFCProxy2022}.
Each of these technologies addresses specific requirements for ensuring efficient, adaptive, and secure communication in both MSI and MSL.
SRv6 enables the dynamic establishment of service paths by embedding routing information directly into packet headers, ensuring that traffic traverses the designated sequence of split sub-models, i.e., Neural Service Functions (NSFs).
TPROXY allows the interception of TCP traffic without modifying the original packets, enabling seamless integration of NSFs into the service chain while maintaining compatibility with existing applications.
The eBPF-based SFC proxy facilitates SRv6 encapsulation and decapsulation, allowing the integration of SFC-unaware NSFs into the SRv6-enabled network.
Together, these technologies ensure flexible and dynamic service chaining, secure communication, and efficient execution of split sub-models in response to changing network conditions.

This work demonstrates the feasibility of SFC-based MSI/MSL through a prototype implementation and experimental evaluation, highlighting its advantages in training performance, latency reduction, and dynamic path reconfiguration.

The main contributions of this paper are summarized as follows:
\begin{enumerate}
  \item We design an SFC-based MSI/MSL architecture that integrates SFC with MSI/MSL, enabling dynamic path reconfiguration and secure communication.
  \item We implement NSFs as stateful network functions using SRv6, TPROXY, and eBPF-based SFC proxy, ensuring compatibility with existing applications.
  \item We demonstrate the feasibility of the proposed architecture through experiments, highlighting its suitability for real-time MSI applications as well as the adaptability to network conditions.
    We also identify challenges in applying the architecture to MSL, particularly regarding increased training time with larger batch sizes and a higher number of splits.
\end{enumerate}

The rest of this paper is organized as follows.
Section~\ref{sec:Related Work} reviews related work.
Section~\ref{sec:Proposed Scheme} introduces the proposed SFC-based MSL/MSI architecture.
Section~\ref{sec:Evaluation Results} presents the feasibility of the proposed architecture through experiments.
Finally, Section~\ref{sec:Conclusion} concludes this paper and gives future work.
\section{Related Work}
\label{sec:Related Work}
\subsection{Multi-hop Split Inference and Learning}
\label{sec:Multi-hop Split Learning}
Model splitting addresses privacy concerns and resource constraints by dividing a global model into multiple split sub-models, albeit at the cost of increased communication overhead.
SL partitions a global model into two segments, placing one on the client side and the other on the server side, thereby safeguarding data privacy while utilizing the server's computational resources~\cite{vepakommaSplitLearningHealth2018,thapaSplitFedWhenFederated2022,kimBargainingGamePersonalized2023,joshiSplitfedLearningClientSide2021,linEfficientParallelSplit2024}.
However, vanilla SL suffers from inefficiencies due to its sequential training process, where only one client is trained at a time.
To mitigate these issues, approaches like SFL~\cite{thapaSplitFedWhenFederated2022} and PSL~\cite{joshiSplitfedLearningClientSide2021,linEfficientParallelSplit2024} have been developed to enable parallel training of client-side models.
Additionally, techniques such as model and activation compression and pruning have been explored to reduce communication overhead and memory usage~\cite{dengModelCompressionHardware2020}.
Chen et al.\ proposed a loss-based asynchronous training method that updates the client-side model less frequently and incorporates 8-bit quantization to further optimize resource usage~\cite{chenCommunicationComputationReduction2021}.

Despite these advancements, SL and its extensions still face challenges related to excessive memory consumption, particularly as global models continue to grow in size.
Smashed data generated during training forms large tensors, which can significantly strain the memory resources of computing nodes.
To address this, MSL has been introduced as an extension of single-hop SL, enabling the use of multiple intermediate nodes between clients and servers~\cite{tiranaMPSLMultihopParallel2024,linHierarchicalSplitFederated2024,wangHiveMindCellularNative2022}.
This extension offers several advantages: (1) it reduces the computational, memory, and storage demands on individual nodes, (2) it optimizes the utilization of idle GPU resources, and (3) it enhances model confidentiality.
Tirana et al.\ proposed a multi-hop parallel SL (MPSL) architecture to alleviate resource limitations~\cite{tiranaMPSLMultihopParallel2024}.
MPSL adopts a self-loop multi-hop path, where the beginning and ending sub-models are executed on the client side, while intermediate parts are executed on the compute nodes, thereby improving data privacy.
Lin et al.\ introduced a hierarchical SFL (HSFL) architecture that connects multiple tiers, from edge devices to the cloud, with each tier responsible for executing a split sub-model~\cite{linHierarchicalSplitFederated2024}.
Wang et al.\ formulated a model splitting problem for MSL in 5G networks, employing a min-cost graph search approach~\cite{wangHiveMindCellularNative2022}.
However, these existing studies on MSI/MSL assume static communication paths between split sub-models, relying on standard IPv4/IPv6 routing.
To overcome this limitation, our SFC-based MSI/MSL architecture aims to establish dynamic communication paths between split sub-models, adapting to changing network conditions.

\subsection{Service Function Chaining}
\label{sec:Service Function Chaining}
With the advancement of network softwarization and programmability, Network Functions (NFs) have been redefined as software components, enabling their deployment as hardware-independent services.
SFC is a fundamental technology aimed at establishing a service path in which traffic traverses multiple NFs in specific order according to service chain requirements~\cite{rfc7665}.
Existing studies on SFC have primarily focused on two categories: network-level use cases for network control and security~\cite{vantuAcceleratingVirtualNetwork2020,pattaranantakulAchievingTrustworthyService2021,haeberleCachingSFCProxy2022,haraEBPFBasedOrderedProof2025}, and application-level use cases such as video and live streaming services~\cite{farahaniSARENASFCEnabledArchitecture2023,ducheneSRv6PipesEnablingInNetwork2018}.
In the former cases, several path verification methods have been proposed to ensure that traffic adheres to the predefined service order.
These include identity-based ordered multi-signature~\cite{pattaranantakulAchievingTrustworthyService2021} and ordered proof of transit~\cite{haraEBPFBasedOrderedProof2025}.
Haeberle et al.\ proposed an eBPF-based SFC proxy, which enables integration of SFC-unaware VNFs to SFC-based networks\cite{haeberleCachingSFCProxy2022}.

In the latter cases, Farahani et al.\ introduced an SFC-based architecture for video streaming services, comprising VNFs for proxy, caching, and transcoding based on FFmpeg~\cite{farahaniSARENASFCEnabledArchitecture2023}.
Transcoding VNFs operate on a per-chunk basis, converting video streaming data into chunks, which are generally larger than individual packet payloads.
This functionality is conceptually similar to preparation for executing split sub-models in MSI.
Since per-packet processing cannot access the full data at once, SFs for MSI must reconstruct received packets into a byte stream by temporarily buffering them until the full data is assembled.
This buffering mechanism has already been implemented in prior work~\cite{ducheneSRv6PipesEnablingInNetwork2018} using SRv6 and transparent proxy techniques (e.g., TPROXY), although it requires SFC-aware functionalities for operation.
Inspired by these studies, we implement transparent SFs for executing split sub-models to incorporate SRv6-based SFC into MSI, by leveraging SRv6, TPROXY, and eBPF-based SFC proxy.
Furthermore, since MSL requires bidirectional communication for training tasks, the architecture extends the application of SFC to manage return traffic effectively.

\section{SFC-based Multi-hop Split Learning and Inference}
\label{sec:Proposed Scheme}
\begin{figure*}[!t]
  \centering
  \begin{minipage}{0.45\hsize}
    \centering
    \includegraphics[width=\columnwidth]{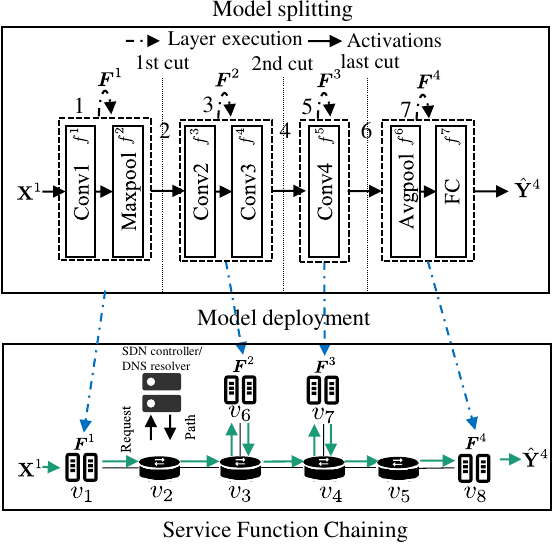}
    \subcaption{SFC-based MSI.}
    \label{fig:system_model:msi}
  \end{minipage}
  \begin{minipage}{0.51\hsize}
    \centering
    \includegraphics[width=\columnwidth]{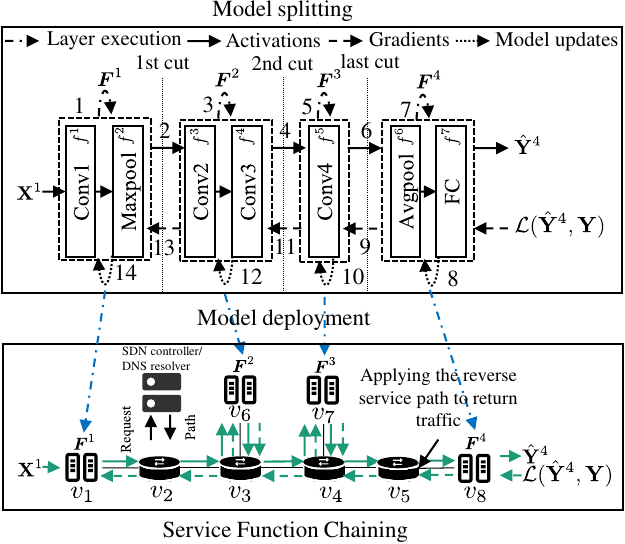}
    \subcaption{SFC-based MSL.}
    \label{fig:system_model:msl}
  \end{minipage}
    \caption{System architecture of SFC-based MSI/MSL.}
    \label{fig:system_model}
\end{figure*}
\begin{table}[!t]
  \caption{Notations.}
  \label{table:notations}
  \small
  \centering
  \newcolumntype{L}{>{\raggedright\arraybackslash}X}
  \begin{tabularx}{\columnwidth}{lL}
    \hline
    Symbol & Description\\
    \hline
    \hline
    $\bm{F}$ & Global model composed of $L$ layers\\
    $L$ & Total number of layers in the global model\\
    $f^l$ & $l$th layer of the global model $\bm{F}$\\
    $\mathbf{X}^{l}$ & Input vector to the $l$th layer\\
    $\hat{\mathbf{Y}}^{l}$ & Output vector (activation) from the $l$th layer\\
    $\mathbf{W}^l$ & Model parameters of the $l$th layer\\
    $\mathbf{G}^l$ & Gradient of the $l$th layer\\
    $\mathbf{Y}$ & Ground truth vector for training\\
    $\mathcal{L}(\hat{\mathbf{Y}}^{L}, \mathbf{Y})$ & Loss function comparing $\hat{\mathbf{Y}}^{L}$ and $\mathbf{Y}$\\
    $\eta$ & Learning rate for gradient descent\\
    $K$ & Number of split sub-models\\
    $\bm{F}^k$ & $k$th sub-model composed of $L^k$ layers (i.e., NSF)\\
    $L^k$ & Number of layers in the $k$th sub-model\\
    $\mathcal{G}=(\V, \E)$ & Substrate network\\
    $\V$ & Set of nodes in the substrate network\\
    $\E$ & Set of links in the substrate network\\
    $M$  & Length of the segment list in SRv6\\
    $m$  & Segments left field in SRv6\\
    \hline
  \end{tabularx}
\end{table}
This section introduces the SFC-based MSI/MSL architecture.
Section~\ref{sec:Fundamental Technologies} explains the fundamental technologies (SRv6, TPROXY, and eBPF-based SFC proxy) utilized in the proposed architecture.
Section~\ref{sec:System Model} describes the system model of the SFC-based MSL/MSI.
Section~\ref{sec:Neural Service Function} details NSFs, which are the core components of the architecture.
Fig.~\ref{fig:system_model} illustrates the system architecture of SFC-based MSI/MSL, and Table~\ref{table:notations} summarizes the notations used throughout this paper.

\subsection{Fundamental Technologies}
\label{sec:Fundamental Technologies}
\subsubsection{Segment Routing over IPv6 (SRv6)}
\label{sec:SRv6}
SRv6 is a source routing protocol that enables packets to follow predefined transit policies~\cite{rfc8986,ventreSegmentRoutingComprehensive2021}.
Nodes in an SRv6 network can act as SR source nodes, SR segment endpoint nodes, or transit nodes.
The SR source node encapsulates packets with a Segment Routing Header (SRH), which contains a segment list and a segments left field.
The segment list is a sequence of $M$ $(M \geq 1)$ Segment Identifiers (SIDs), representing nodes to be visited, listed in reverse order of traversal.
The segment left field indicates the next active SID.
SR segment endpoint nodes process the SRH, updating the destination address and forwarding the packet accordingly.
Transit nodes simply forward packets based on standard IPv6 routing.

\subsubsection{Transparent Proxy (TPROXY)}
\label{sec:Transparent Proxy}
\begin{figure}[!t]
 \centering
\begin{lstlisting}[style=bash]
ip6tables -t mangle -N DIVERT
ip6tables -t mangle -A DIVERT -p tcp -d ${TARGET_IP} --dport ${TARGET_PORT} -j MARK --set-mark ${FWMARK}
ip6tables -t mangle -A DIVERT -p tcp -d ${TARGET_IP} --dport ${TARGET_PORT} -j TPROXY --on-port ${TPROXY_PORT} --tproxy-mark ${FWMARK}/${FWMARK}
ip6tables -t mangle -A DIVERT -j ACCEPT
ip6tables -t mangle -A PREROUTING -p tcp -d ${TARGET_IP} --DPORT ${TARGET_PORT} -j DIVERT
ip -6 route add local ::/0 dev lo table ${FWMARK}
\end{lstlisting}
  \caption{Example configuration for TPROXY.}
 \label{fig:tproxy_configuration}
\end{figure}
TPROXY is a Linux kernel module that intercepts TCP traffic without altering the original packets~\cite{tproxy}.
It is typically configured using iptables~\cite{iptables}.
When a packet matches specific rules, it is marked and redirected to a predefined target port.
This allows clients to communicate via proxies while perceiving a direct connection to the server.
Fig.~\ref{fig:tproxy_configuration} shows an example configuration for TPROXY, where \texttt{\$\{TARGET\_IP\}} and \texttt{\$\{TARGET\_PORT\}} denote the IP address and port number of the target server, respectively.
\texttt{\$\{TPROXY\_PORT\}} denotes the port number of the TPROXY server.
\texttt{\$\{FWMARK\}} denotes a mark to identify the intercepted packets.
In this configuration, when a received packet's destination address and port number match \texttt{\$\{TARGET\_IP\}} and \texttt{\$\{TARGET\_PORT\}}, the packet is marked with \texttt{\$\{FWMARK\}} and redirected to the TPROXY server listening on \texttt{\$\{TPROXY\_PORT\}}.
The TPROXY server then receives the intercepted packet, processes it according to the application logic, and subsequently forwards it to the target server by establishing a new connection.
This indicates that the connection between the client and the TPROXY server and that between the TPROXY server and the target server are established independently.

\subsubsection{extended Berkeley Packet Filter (eBPF) based SFC Proxy}
\label{sec:extended Berkeley Packet Filter}
eBPF is a virtual machine within the Linux kernel that allows user-defined programs to run safely and efficiently~\cite{soldaniEBPFNewApproach2023,sharafExtendedBerkeleyPacket2022}.
eBPF programs can be attached to Traffic Control (TC) hooks to process packets at the ingress or egress points.
BPF maps serve as key-value stores for sharing data between eBPF programs and user-space applications.
Thanks to its portability and flexibility, eBPF has been applied to various network contexts, including packet filtering~\cite{choeEBPFXDPBased2020}, SRv6~\cite{xhonneuxLeveragingEBPFProgrammable2018}, SFC proxies~\cite{haeberleCachingSFCProxy2022,haraEBPFBasedOrderedProof2025}, and VNFs~\cite{vantuAcceleratingVirtualNetwork2020, haraPracticalityInKernelUserspace2024}.
In this paper, eBPF is used to implement an SFC proxy that performs SRv6 encapsulation and decapsulation, enabling the integration of SFC-unaware VNFs into an SRv6-enabled network~\cite{haeberleCachingSFCProxy2022,haraEBPFBasedOrderedProof2025}.

\subsection{System Model}
\label{sec:System Model}
\subsubsection{Global Model}
\label{sec:Global Model}
The global model $\bm{F}=(f^1, \ldots, f^L)$ is represented as a sequence of $L$ layers, where the $l$th layer $f^l$ contains model parameters $\mathbf{W}^l$.
In the \emph{inference} process, only forward propagation is performed, as illustrated in Fig.~\ref{fig:system_model:msi}, whereas the \emph{training/learning} process involves both forward propagation and backward propagation, as shown in Fig.~\ref{fig:system_model:msl}.

During forward propagation, the global model $\bm{F}$ takes an input vector $\mathbf{X}^{1}$ and produces an output vector $\hat{\mathbf{Y}}^{L}$, expressed as $\hat{\mathbf{Y}}^{L}=\bm{F}(\mathbf{X}^{1})$.
In this process, each layer $f^l$ computes its output vector $\hat{\mathbf{Y}}^{l}$ from the input vector $\mathbf{X}^{l}$ and passes it to the subsequent layer as $\mathbf{X}^{l+1} = \hat{\mathbf{Y}}^{l}$ for $l=1, \ldots, L-1$.
For inference, $\hat{\mathbf{Y}}^{L}$ represents the inference results, while for training, it is used to evaluate a loss function $\mathcal{L}(\hat{\mathbf{Y}}^{L}, \mathbf{Y})$, which quantifies the difference between $\hat{\mathbf{Y}}^{L}$ and the ground truth vector $\mathbf{Y}$.
In backward propagation, the gradient $\mathbf{G}=(\mathbf{G}^{1}, \ldots, \mathbf{G}^{L})$ of the loss function $\mathcal{L}$ is calculated with respect to the model parameters $\mathbf{W}=(\mathbf{W}^1, \ldots, \mathbf{W}^L)$.
The model parameters are then updated as $\mathbf{W}^l \leftarrow \mathbf{W}^l - \eta \mathbf{G}^l$ for each layer $l$, where $\eta$ denotes the learning rate,

\subsubsection{Model Splitting}
\label{sec:Model Splitting}
The global model $\bm{F}$ is disjointly partitioned into $K$ sub-models $\bm{F}^k$ for $k=1, \ldots, K$ (i.e., $\bm{F}=(\bm{F}^1, \ldots, \bm{F}^K)$), where $2 \leq K \leq L$.
Each sub-model $\bm{F}^k$ consists of $L^k$ sequential layers, expressed as $\bm{F}^k=(f^l, \ldots, f^{l+L^k-1})$, with $L=\sum_{k=1}^K L^k$.
In the upper layer of Fig.~\ref{fig:system_model}, the global model with $L=7$ layers is divided into $K=4$ sub-models: $\bm{F}^1$, $\bm{F}^2$, $\bm{F}^3$, and $\bm{F}^4$, containing $L^1=2$, $L^2=2$, $L^3=1$, and $L^4=2$ layers, respectively.

During forward propagation (steps 1--7 in Figs.~\ref{fig:system_model:msi} and \ref{fig:system_model:msl}), the overall latency comprises two components: forward propagation time and activation transmission time.
The forward propagation time refers to the computing time required for each sub-model to perform layer execution from input vectors (steps 1, 3, 5, and 7 in Figs.~\ref{fig:system_model:msi} and \ref{fig:system_model:msl}), while the activation transmission time denotes the time required to transmit activations between sub-models over the communication network (steps 2, 4, and 6 in Figs.~\ref{fig:system_model:msi} and \ref{fig:system_model:msl}).
For instance, the first sub-model $\bm{F}^1$ computes the activation $\hat{\mathbf{Y}}^{L^1}$ from the input vector $\mathbf{X}^{1}$ (step 1 in Figs.~\ref{fig:system_model:msi} and \ref{fig:system_model:msl}) and transmits it to the next sub-model $\bm{F}^2$ (step 2 in Figs.~\ref{fig:system_model:msi} and \ref{fig:system_model:msl}).
Each intermediate sub-model $\bm{F}^k$ ($k = 2 \ldots, K-1$) receives the activation $\hat{\mathbf{Y}}^{l}$ from the preceding sub-model $\bm{F}^{k-1}$ (steps 2 and 4 in Figs.~\ref{fig:system_model:msi} and \ref{fig:system_model:msl}), computes the activation $\hat{\mathbf{Y}}^{l+L^k-1}$ (steps 3 and 5 in Figs.~\ref{fig:system_model:msi} and \ref{fig:system_model:msl}), and forwards it to the subsequent sub-model $\bm{F}^{k+1}$ through the communication network (steps 4 and 6 in Figs.~\ref{fig:system_model:msi} and \ref{fig:system_model:msl}).
Finally, the last sub-model $\bm{F}^K$ computes the output vector $\hat{\mathbf{Y}}^{L}$ (step 7 in Figs.~\ref{fig:system_model:msi} and \ref{fig:system_model:msl}).

In backward propagation (steps 8--14 in Fig.~\ref{fig:system_model:msl}), the overall latency similarly consists of backward propagation time and gradient transmission time.
The backward propagation time refers to the computation time required for each sub-model to calculate gradients and update model parameters (steps 8, 10, 12, and 14 in Fig.~\ref{fig:system_model:msl}), while the gradient transmission time accounts for the time needed to transmit gradients between sub-models over the network (steps 9, 11, and 13 in Fig.~\ref{fig:system_model:msl}).
For instance, the last sub-model $\bm{F}^K$ computes the gradients $(\mathbf{G}^{L-L^K+1}, \ldots, \mathbf{G}^L)$, updates the model parameters $(\mathbf{W}^{L-L^K+1}, \ldots, \mathbf{W}^L)$ (step 8 in Fig.~\ref{fig:system_model:msl}), and transmits the gradient $\mathbf{G}^{L-L^K+1}$ to the preceding sub-model $\bm{F}^{K-1}$ (step 9 in Fig.~\ref{fig:system_model:msl}).
Each intermediate sub-model $\bm{F}^k$ ($k = K-1, \ldots, 2$) receives the gradient $\mathbf{G}^{l+L^k}$ from the subsequent sub-model $\bm{F}^{k+1}$ (steps 9 and 11 in Fig.~\ref{fig:system_model:msl}), computes the gradients $(\mathbf{G}^{l}, \ldots, \mathbf{G}^{l+L^k-1})$, updates the model parameters $(\mathbf{W}^{l}, \ldots, \mathbf{W}^{l+L^k-1})$ (steps 10 and 12 in Fig.~\ref{fig:system_model:msl}), and transmits the gradient $\mathbf{G}^{l}$ to the preceding sub-model $\bm{F}^{k-1}$ (steps 11 and 13 in Fig.~\ref{fig:system_model:msl}).
Finally, the first sub-model $\bm{F}^1$ receives the gradient $\mathbf{G}^{L^1}$ (step 13 in Fig.~\ref{fig:system_model:msl}) and updates the model parameters $(\mathbf{W}^{1}, \ldots, \mathbf{W}^{L^1})$ (step 14 in Fig.~\ref{fig:system_model:msl}).

Thus, the sequential execution of the sub-models $\bm{F}^1, \ldots, \bm{F}^K$ corresponds to the execution of the global model.
As the number of splitting points increases (i.e., as $K$ increases), the communication overhead grows due to more frequent exchanges of activations and gradients.
This leads to longer transmission times for both activations and gradients.
During training, each sub-model must wait for the subsequent sub-model to return gradients after completing its forward propagation, introducing a delay referred to as the waiting time.
In contrast, during inference, this waiting time is avoided when sub-models do not require downstream outputs, such as inference results.

\subsubsection{Substrate Network}
\label{sec:Substrate Network}
Consider a substrate network represented as $\mathcal{G}=(\V, \E)$, where $\V$ denotes the set of nodes and $\E$ represents the set of links.
Each node $v \in \V$ is equipped with computing resources $C_v^\mathrm{com}$, memory resources $C_v^\mathrm{mem}$, and storage resources $C_v^\mathrm{sto}$.
Similarly, each link $e \in \E$ is characterized by its bandwidth $B_e$.
The substrate network is assumed to support SRv6 and Software Resolved Networks (SRNs)~\cite{lebrunSoftwareResolvedNetworks2018}.
An SDN controller or DNS resolver is responsible for managing NSFs deployed on computing nodes and providing clients with a service path from an origin node to a destination node based on their requests.

\subsubsection{Service Chain}
A service chain is defined as an ordered sequence of $K$ NSFs, where each NSF $\bm{F}^k$ is responsible for executing the $k$th split sub-model.
The detailed NSF functionalities are elaborated in Section~\ref{sec:Neural Service Function}.
In Fig.~\ref{fig:system_model}, the four sub-models, $\bm{F}^1$, $\bm{F}^2$, $\bm{F}^3$, and $\bm{F}^4$, are deployed as NSFs on computing nodes $v_1$, $v_6$, $v_7$, and $v_8$, respectively.

\subsubsection{TCP Proxy Chaining Mechanisms}
\label{sec:TCP Proxy Chaining Mechanisms}
To enforce traffic traversal through designated NSFs, three types of chaining mechanisms are considered: (1) traditional TCP proxy chaining over IPv4/IPv6, (2) transparent TCP proxy chaining over IPv4/IPv6, and (3) transparent TCP proxy chaining over SRv6.

Traditional TCP proxy chaining, as adopted in existing MSL/MSI studies~\cite{tiranaMPSLMultihopParallel2024}, establishes a chained TCP flow by creating connections between the client and the first proxy, and between each subsequent proxy in the chain, until the last proxy forwards the traffic to the target server.
All packets from the client traverse the proxies in the service chain based on standard IPv4/IPv6 routing rules.
However, this approach requires clients to be aware of the first NSF in the service chain to establish service paths.
For instance, if the service chain is $(\bm{F}^1, \bm{F}^2, \bm{F}^3, \bm{F}^4)$, the resulting service path is $v_1 \rightarrow v_2 \rightarrow v_3 \rightarrow v_6 \rightarrow v_3 \rightarrow v_4 \rightarrow v_7 \rightarrow v_4 \rightarrow v_5 \rightarrow v_8$.

In transparent TCP proxy chaining, clients are only aware of the target server, while intermediate proxies intercept traffic without modifying the original packets by leveraging TPROXY~\cite{tproxy}.
However, if the routing rules are not properly configured, this approach may fail to meet service chain requirements, as traffic might bypass intermediate proxies.
For example, in Fig.~\ref{fig:system_model}, if TPROXY configuration is applied to nodes $v_6$ and $v_7$, the executions of sub-models $\bm{F}^2$ and $\bm{F}^3$ may be skipped.

To ensure traffic traverses TPROXY servers, the transparent TCP proxy chaining over SRv6 (i.e., the proposed architecture) combines SRv6 and TPROXY to establish service paths.
This guarantees that traffic from a client $v_1$ to the target server $v_8$ always traverses the designated TPROXY on nodes $v_6$ and $v_7$.
For instance, if the service chain is $(\bm{F}^1, \bm{F}^2, \bm{F}^3, \bm{F}^4)$, the resulting service path is $v_1 \rightarrow v_2 \rightarrow v_3 \rightarrow v_6 \rightarrow v_3 \rightarrow v_4 \rightarrow v_7 \rightarrow v_4 \rightarrow v_5 \rightarrow v_8$ during forward propagation.
For MSL, traffic control for backward propagation must also be addressed.
A straightforward method is to enforce the backward propagation traffic to follow the reverse order of the service chain $(\bm{F}^4, \bm{F}^3, \bm{F}^2, \bm{F}^1)$.
In this case, the backward propagation traffic retraces the reversed service path of the forward propagation as a response to the client.
The resulting service path is $v_8 \rightarrow v_5 \rightarrow v_4 \rightarrow v_7 \rightarrow v_4 \rightarrow v_3 \rightarrow v_6 \rightarrow v_3 \rightarrow v_2 \rightarrow v_1$.
The designated service path can dynamically adapt based on communication requirements and network conditions by manipulating the segment list.

\subsection{Neural Service Function (NSF)}
\label{sec:Neural Service Function}
\begin{figure*}[!t]
  \centering
    \includegraphics[width=0.8\textwidth]{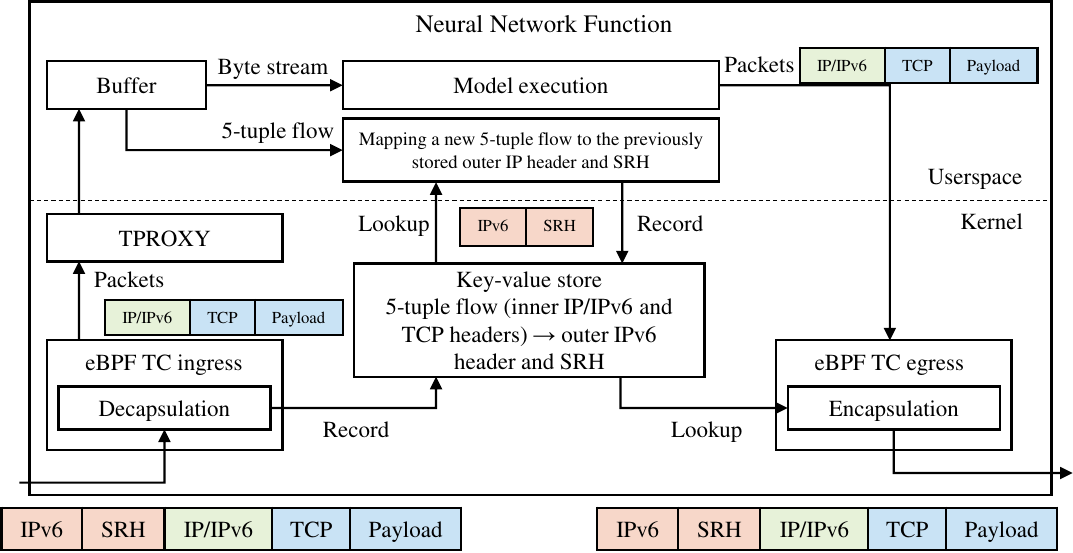}
    \caption{Architecture of a neural service function (NSF).}
    \label{fig:neural_service_function}
\end{figure*}
\subsubsection{Overview}
An NSF is implemented as a stateful network function responsible for executing a split sub-model.
Fig.~\ref{fig:neural_service_function} depicts the architecture of the NSF.
Operating as a TPROXY server over SRv6, the NSF enables seamless integration into a service chain without requiring modifications to client-side or server-side applications.
In this configuration, the client communicates with the server without being aware of the intermediate NSFs, while traffic transparently traverses the designated NSFs in the specified order.

Upon receiving an incoming packet, the NSF performs SRv6 decapsulation using eBPF-based SFC proxy and evaluates the packet against predefined TPROXY rules.
If the packet matches the TPROXY rules, it is redirected to the NSF's designated target port.
The NSF buffers incoming packets until the full data is reconstructed as a byte stream, enabling the execution of the split sub-model.
After executing the sub-model, the NSF generates intermediate activations and forwards them to the next NSF as packets encapsulated with an SRH using the eBPF-based SFC proxy.
In the case of training, the NSF waits for gradients from the subsequent NSF, performs backward propagation, and transmits the computed gradients to the preceding NSF.

\subsubsection{SRv6 Packet}
\label{sec:SRv6 Packet}
\begin{figure}[!t]
  \centering
    \includegraphics[width=\columnwidth]{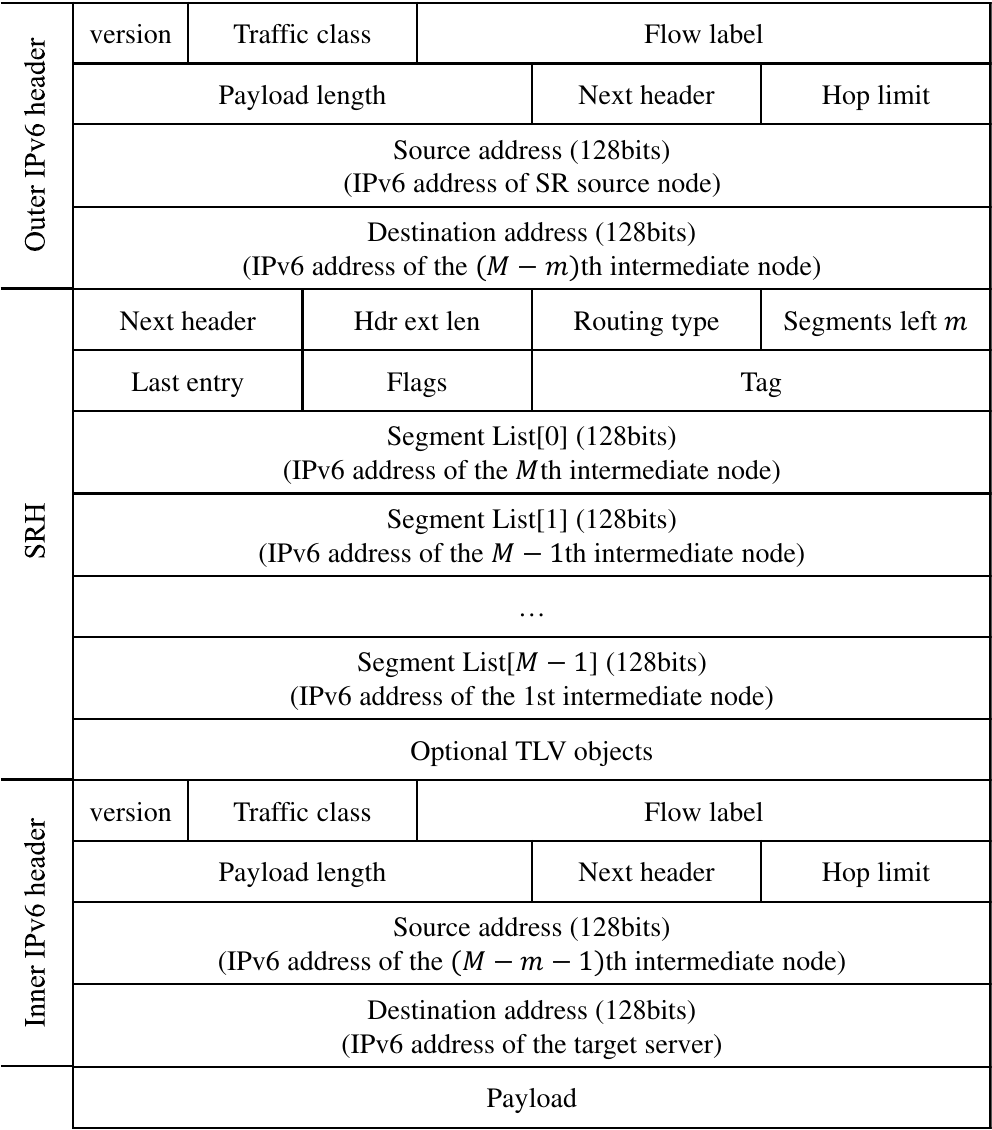}
    \caption{Structure of an SRv6 packet with the segments left field set to $m$.}
    \label{fig:srv6_packet}
\end{figure}
SRv6 encapsulation is a critical component of the SFC-based MSL/MSI architecture, enabling the designation of specific service paths.
Each packet sent from clients is encapsulated with an outer IPv6 header and an SRH by an SR source node.
The SDN controller or DNS resolver provides a sequence of SIDs in response to client requests.
Fig.~\ref{fig:srv6_packet} illustrates an SRv6 packet when the segments left field is set to $m$.
In the SRH, the segment list field contains $M$ SIDs, and the segments left field indicates $m$.
Consequently, the destination address field of the outer IPv6 header holds the $(M-m)$th SID.
The source address field of the outer IPv6 header always holds the IPv6 address of the SR source node.
Since the packet originates from the $(M-m-1)$th SR endpoint node, the source address field of the inner IPv6 header is set to the IP address of that node.
The destination address of the inner IPv6 header consistently holds the target server's IPv6 address, as enforced by the TPROXY configuration.
Although Fig.~\ref{fig:srv6_packet} assumes the inner packet header uses IPv6, IPv4 is also supported.

\subsubsection{SFC Proxy for SRv6 Encapsulation and Decapsulation}
\label{sec:SFC Proxy}
The SFC proxy for SRv6 encapsulation and decapsulation is implemented using eBPF.
The eBPF program for SRv6 encapsulation is attached to the TC egress hook, while the program for decapsulation is attached to the TC ingress hook.

The eBPF program at the TC ingress inspects incoming packets to determine if the destination address in the outer IPv6 header matches the NSF's SID.
If a match is found, the program decapsulates the SRv6 packet, updates the fields in the IPv6 header and SRH, and stores the 5-tuple flow of the inner packet as a key in a BPF map, along with the outer IPv6 header and SRH as the associated value.
If no match is found, the packet is forwarded to the next node based on the forwarding policy.

For outgoing packets, the eBPF program at the TC egress inspects the packet and retrieves the corresponding outer IPv6 header and SRH from the BPF map using the 5-tuple flow of the outgoing packet as a key.
If the retrieval is successful, the program encapsulates the outgoing packet with the SRH.

\subsubsection{Transparent TCP Proxy Chaining over SRv6}
\label{sec:Transparent TCP Proxy Chaining over SRv6}
\begin{figure}[!t]
  \centering
    \includegraphics[width=0.9\columnwidth]{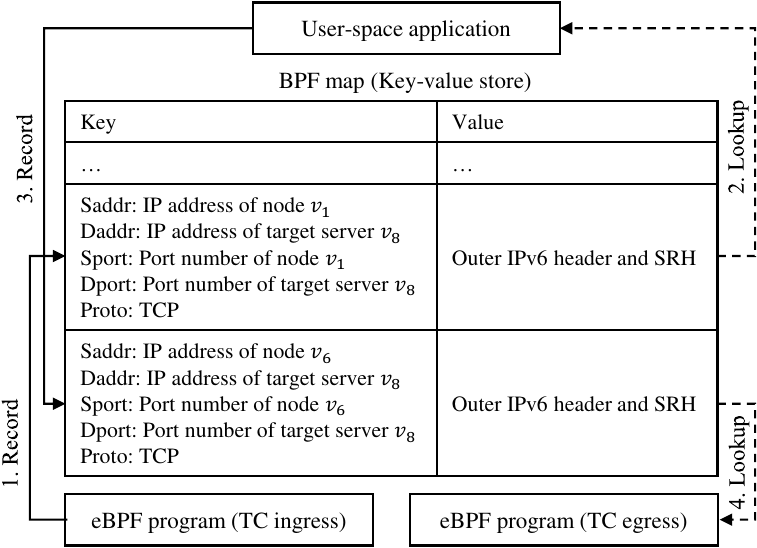}
    \caption{Key-value store updates for transparent TCP proxy chaining over SRv6.}
    \label{fig:bpf_map}
\end{figure}
When an incoming packet satisfies the TPROXY rules, it is intercepted by the NSF.
Therefore, the NSF needs to establish a new TCP connection to the target server in place of the client.
From an SRv6 perspective, this new TCP connection must be associated with the existing TCP connection between the previous NSF (or the client) and the target server.
To this end, the NSF first retrieves the outer IPv6 header and SRH from the BPF map using the 5-tuple flow of the original incoming packet as a key, and then records the 5-tuple flow of the new TCP connection as a key and the retrieved outer IPv6 header and SRH as a value into the BPF map.
This series of processes realizes transparent TCP proxy chaining over SRv6.

Fig.~\ref{fig:bpf_map} depicts an example of key-value store updates for transparent TCP proxy chaining over SRv6, assuming that node $v_6$ intercepts an SRv6 packet from node $v_1$ and the packet traverses the predefined service path in Fig.~\ref{fig:system_model}.
As mentioned in Section~\ref{sec:SFC Proxy}, the 5-tuple flow, outer IPv6 header, and SRH of the received packet are stored to the BPF map (step 1 in Fig.~\ref{fig:bpf_map}).
Focusing on the components of the 5-tuple flow, we observe from Fig.~\ref{fig:srv6_packet} that the protocol number is 6 (i.e., TCP), the source address is the IP address of node $v_1$, the destination address is the IP address of the target server (i.e., node $v_8$), the source port number corresponds to $v_1$'s port number, and the destination port number corresponds to the target server's port number.
The NSF retrieves the outer IPv6 header and the SRH from the BPF map using this 5-tuple flow as a key (step 2 in Fig.~\ref{fig:bpf_map}).
After the execution of the split sub-model, it attempts to send the activations to the target server (i.e., node $v_8$).
Before sending the activations to the target server, the NSF records the 5-tuple flow of the new TCP connection as a key and the retrieved outer IPv6 header and SRH as a value into the BPF map (step 3 in Fig.~\ref{fig:bpf_map}).
The eBPF-based SFC proxy performs SRv6 encapsulation of outgoing packets using the stored outer IPv6 header and SRH (step 4 in Fig.~\ref{fig:bpf_map}).
This ensures that packets sent from node $v_6$ with the destination address set to the IP address of node $v_8$ are consistently intercepted by node $v_7$, enabling seamless traversal through the designated service chain.

\subsubsection{Execution of Sub-Model}
\label{sec:Execution of Sub-Model}
Due to the size limitations of Ethernet frames (1,500 bytes) and jumbo Ethernet frames (9,000 bytes), packet processing cannot access the entire data at once.
To address this, NSFs buffer all received packets and reconstruct the complete data as a byte stream before executing the assigned sub-model.
Using the reconstructed byte stream as input, each NSF processes the data through its sub-model and generates the output (i.e., activations), which is subsequently forwarded to the next NSF in the chain.

\subsubsection{Return Traffic}
\label{sec:Return traffic}
The training process requires backward propagation (return traffic) in addition to forward propagation.
After receving the activation from the preceding NSF, the target server calculates the gradients, updates its model parameters, and sends the gradients back to the preceding NSF.
The return traffic from the target server to the client is encapsulated with an SRH, where the segment list is reversed based on the SRH received from the client or previous NSFs.
Upon receiving the response, each intermediate NSF decapsulates the SRv6 packets, reconstructs the inner packets into a byte stream, computes the gradients, updates the model parameters, and forwards the gradients to the preceding NSF as packets encapsulated with the updated SRH.

\section{Evaluation Results}
\label{sec:Evaluation Results}
\subsection{Experiment Setup}
\begin{figure}[!t]
  \centering
    \includegraphics[width=0.9\columnwidth]{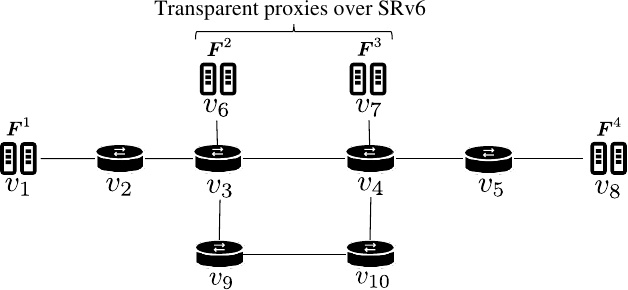}
    \caption{Experimental setup and network topology.}
    \label{fig:evaluation_environment}
\end{figure}
The prototype of the proposed architecture is implemented using C and C++ with the Libtorch~\cite{paszkePyTorchImperativeStyle2019} and libbpf~\cite{libbpf} libraries.
The experiments are conducted on a server equipped with a 36-core Intel(R) Core(TM) i9-10980XE CPU, a 128\,GB RAM, an NVIDIA RTX A6000 GPU, and Ubuntu~22.04 LTS (kernel version 6.2.0).
The network topology, as shown in Fig.~\ref{fig:evaluation_environment}, is emulated using Mininet~\cite{mininet} and supports jumbo Ethernet frames.
Node $v_1$ functions as the client, executing the first sub-model $\bm{F}^1$ and initiating requests for the execution of the remaining sub-models on the target server $v_8$.
Outgoing packets from $v_1$ are encapsulated with an SRH at $v_2$, where the segment list and segments left fields are initialized to $[v_6, v_7, v_5]$ and 3, respectively.
Nodes $v_6$ and $v_7$ serve as NSFs, executing the intermediate sub-models $\bm{F}^2$ and $\bm{F}^3$, respectively.
TPROXY rules are configured using iptables (version 1.8.7) and applied to $v_6$ and $v_7$.
The eBPF-based SFC proxies, implemented with the libbpf library, are attached to both the TC ingress and egress hooks on $v_6$ and $v_7$.
Node $v_5$ acts as an SR segment endpoint node, decapsulating SRv6 packets and forwarding the inner packets to $v_8$ based on the forwarding/routing table.
Node $v_8$ functions as the target server, executing the final sub-model $\bm{F}^4$ and returning either inference results (in the case of inference) or gradients to the preceding NSF (in the case of training).
The service path is defined as $v_1 \rightarrow v_2 \rightarrow v_3 \rightarrow v_6 \rightarrow v_3 \rightarrow v_4 \rightarrow v_7 \rightarrow v_4 \rightarrow v_5 \rightarrow v_8$.
The default bandwidth $B_e$ for each link $e$ is set to 1,000\,Mbps using TC for both inference and training.

\begin{table}[!t]
  \caption{Specification of split sub-models used in the evaluation ($b=1$).}
  \label{table:spec_of_split_model}
  \centering
  \newcolumntype{L}{>{\raggedright\arraybackslash}X}
  \newcolumntype{R}{>{\raggedleft\arraybackslash}X}
  \begin{tabularx}{\columnwidth}{lRR}
    \hline
    Split sub-model & Input dimension ($b=1$)& Output dimension ($b=1$)\\
    \hline
    \hline
    $\bm{F}^1$ ($L^1=2$)  & 3,072 & 4,096\\
    $\bm{F}^2$ ($L^2=8$)  & 4,096 & 2,048\\
    $\bm{F}^3$ ($L^3=9$)  & 2,048 & 1,024\\
    $\bm{F}^4$ ($L^4=18$) & 1,024 & 100\\
    \hline
  \end{tabularx}
\end{table}

Considering model suitability for inference tasks, we use a reduced-dimension version of ResNet101 as the global model, where the output channel size is scaled down to one-fourth of the original configuration~\cite{heDeepResidualLearning2016}.
To balance sub-models' modularity and granularity, we treat a building block as a unit of model splitting and therefore set $L=37$.
The global model is split into four sub-models: $\bm{F}^1$, $\bm{F}^2$, $\bm{F}^3$, and $\bm{F}^4$, containing $L^1=2$, $L^2=8$, $L^3=9$, and $L^4=18$ layers, respectively.
These sub-models are deployed on $v_1$, $v_6$, $v_7$, and $v_8$.
The training process uses a Stochastic Gradient Descent (SGD) optimizer with an initial learning rate $\eta=0.1$, momentum of 0.9, and weight decay of $5\times10^{-4}$.
The learning rate decreases by a factor of 5 after the 60th, 120th, and 160th epochs~\cite{devriesImprovedRegularizationConvolutional2017}.
The CIFAR-100 dataset~\cite{krizhevskyLearningMultipleLayers2009}, consisting of 50,000 training images and 10,000 testing images across 100 classes, is used.
The mini-batch size is set to $b=128$ for training, and Table~\ref{table:spec_of_split_model} outlines the input and output dimensions of each split sub-model when $b=1$.

Evaluation metrics include per-round inference latency and testing accuracy for inference, as well as per-round training latency, training accuracy, and training loss for training.
Here, one round refers to the completion of inference or training by all sub-models for a single mini-batch containing $b$ data samples.
Per-round inference latency measures the time required for one round of inference, including forward propagation and activation transmission.
It is noted that the activation transmission time is divided into two components: the sending time and the receiving time.
The sending time refers to the duration from the initiation to the completion of sending activations, while the receiving time is the duration from the start to the end of receiving them.
Per-round training latency measures the time required to complete one round of training, including forward propagation, activation transmission, backward propagation, gradient transmission, and waiting time.
Similarly, the gradient transmission time is divided into the sending time and the receiving time.
The sending time refers to the duration from the initiation to the completion of sending gradients, while the receiving time is the duration from the start to the end of receiving them.
The waiting time is defined as the elapsed time from the transmission of the activations to the reception of corresponding gradients.
Testing accuracy is defined as the ratio of correctly classified samples to all samples.
Training accuracy is defined as the ratio of correctly classified samples per mini-batch, while training loss is calculated using a cross-entropy loss function.
The first iteration is excluded from latency evaluation to eliminate initialization and caching effects.
The number of epochs is set to 200, where one epoch refers to processing the entire dataset.

As a baseline, the existing MSL/MSI architecture is implemented using traditional TCP proxy chaining.
In this approach, the client $v_1$ sends requests to the first proxy $v_6$, which forwards them to the next proxy $v_7$, and finally to the target server $v_8$.

\subsection{Training and Testing Accuracy}
\label{sec:Training accuracy}
\begin{figure}[!t]
  \centering
  \begin{minipage}{\hsize}
    \centering
    \includegraphics[width=\columnwidth]{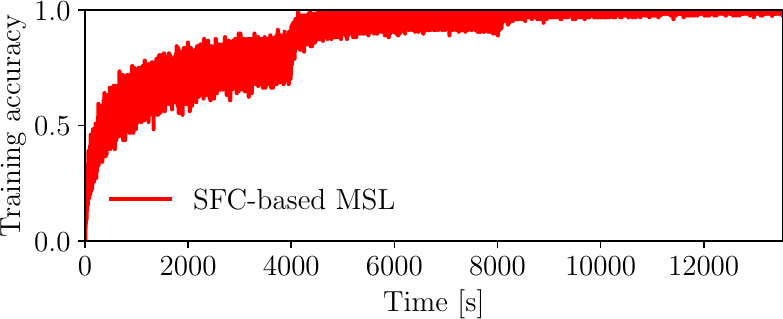}
    \subcaption{Proposed SFC-based MSL.}
    \label{fig:training_accuracy:proposal}
  \end{minipage}
  \begin{minipage}{\hsize}
    \centering
    \includegraphics[width=\columnwidth]{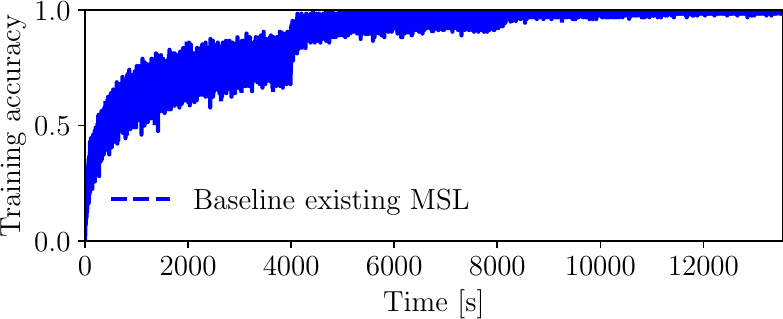}
    \subcaption{Baseline existing MSL.}
    \label{fig:training_accuracy:existing}
  \end{minipage}
  \caption{Training accuracy over time.}
  \label{fig:training_accuracy}
\end{figure}
\begin{figure}[!t]
  \centering
  \begin{minipage}{\hsize}
    \centering
    \includegraphics[width=\columnwidth]{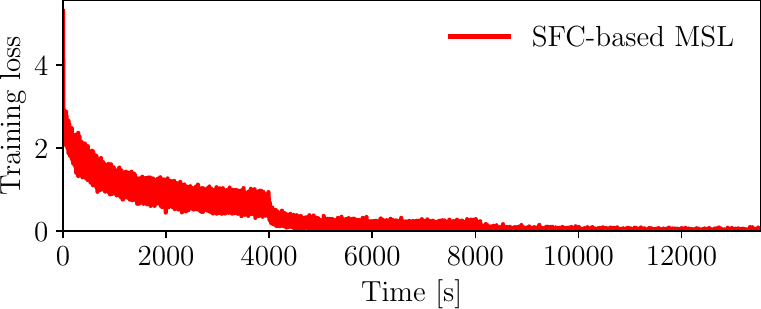}
    \subcaption{Proposed SFC-based MSL.}
    \label{fig:training_loss:proposal}
  \end{minipage}
  \begin{minipage}{\hsize}
    \centering
    \includegraphics[width=\columnwidth]{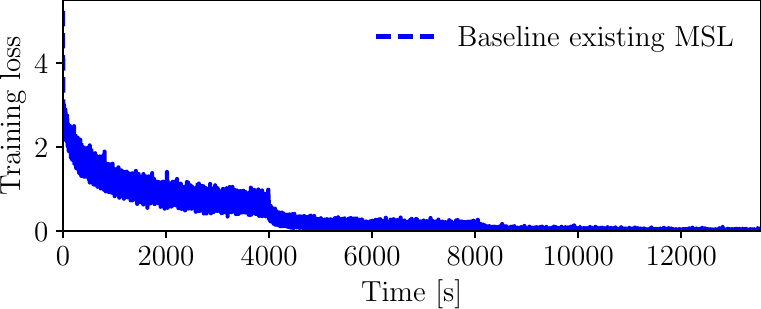}
    \subcaption{Baseline existing MSL.}
    \label{fig:training_loss:existing}
  \end{minipage}
  \caption{Training loss over time.}
  \label{fig:training_loss}
\end{figure}
Figs.~\ref{fig:training_accuracy} and \ref{fig:training_loss} illustrate the training accuracy and loss over time, respectively.
The horizontal axis represents time in seconds rather than epochs.
Both the proposed and existing architectures exhibit similar trends, with accuracy improving and loss decreasing as training progresses.
After approximately 13,000 seconds, both architectures complete the 200th training epoch.
By the 200th epoch, the training accuracy reaches 99.8\% for both architectures, while the training loss is 0.002 for both.
These results confirm that the NSFs effectively train their respective sub-models.
The testing accuracy reaches 71.1\% for both architectures after 200 epochs.

\begin{figure*}[!t]
  \centering
  \begin{minipage}{0.49\hsize}
    \centering
    \includegraphics[width=\columnwidth]{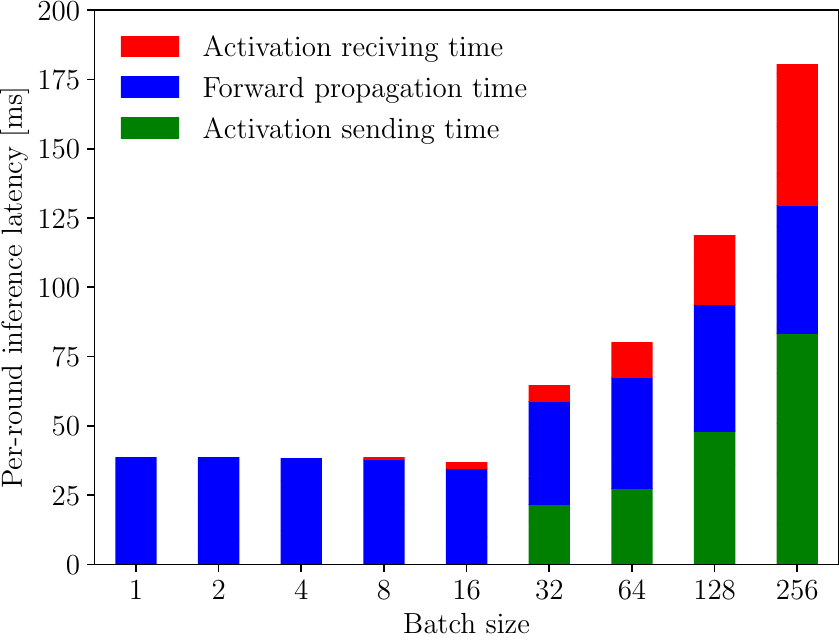}
    \subcaption{Proposed SFC-based MSI.}
    \label{fig:impact_of_bs_on_inference_latency:proposal}
  \end{minipage}
  \begin{minipage}{0.49\hsize}
    \centering
    \includegraphics[width=\columnwidth]{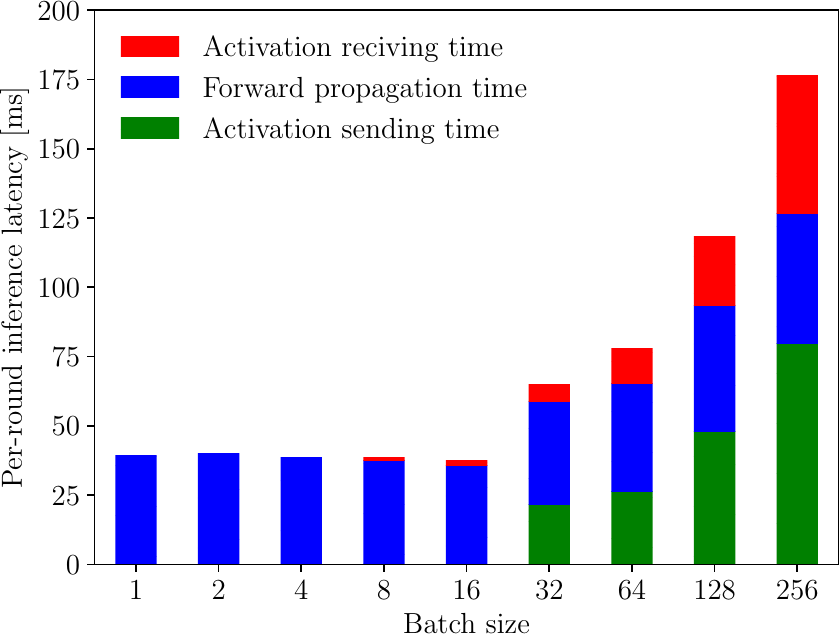}
    \subcaption{Baseline existing MSI.}
    \label{fig:impact_of_bs_on_inference_latency:existing}
  \end{minipage}
  \caption{Comparison of the impact of batch size on per-round inference latencies between the proposed SFC-based MSI and the baseline existing MSI.}
  \label{fig:impact_of_bs_on_inference_latency}
\end{figure*}
\subsection{Inference Latency}
\label{sec:Inference Latency}
This section evaluates the inference latency of the proposed SFC-based MSI and the baseline existing MSI.
Figs.~\ref{fig:impact_of_bs_on_inference_latency:proposal} and \ref{fig:impact_of_bs_on_inference_latency:existing} illustrate the impact of batch size on per-round inference latencies for both architectures.
Both architectures exhibit similar trends: the forward propagation time remains nearly constant due to GPU acceleration and I/O overhead, while the activation transmission time increases as the batch size grows.
The results indicate no significant difference in inference latencies between the proposed SFC-based MSI and the baseline existing MSI.
This similarity arises because the overhead of SRv6 encapsulation and decapsulation is insignificant compared to the computational workload of forward propagation.
Specifically, the per-round inference latency is 38.7\,ms for the proposed SFC-based MSI and 39.2\,ms for the baseline existing MSI when $b=1$.
The primary bottleneck is forward propagation, as the communication latency is minimal due to the small size of smashed data when the mini-batch size is set to $b=1$.
As the batch size increases, the bottleneck shifts from forward propagation to activation transmission.

This observation suggests that applying SFC to MSI is particularly effective in scenarios where real-time performance is critical, such as when the mini-batch size is set to $b=1$.
In such cases, the communication overhead is minimal due to the small size of the smashed data, ensuring that the latency introduced by SRv6 encapsulation and decapsulation remains negligible.
This makes the proposed SFC-based MSI architecture well-suited for real-time inference tasks.

\begin{figure*}[!t]
  \centering
  \begin{minipage}{0.49\hsize}
    \centering
    \includegraphics[width=\columnwidth]{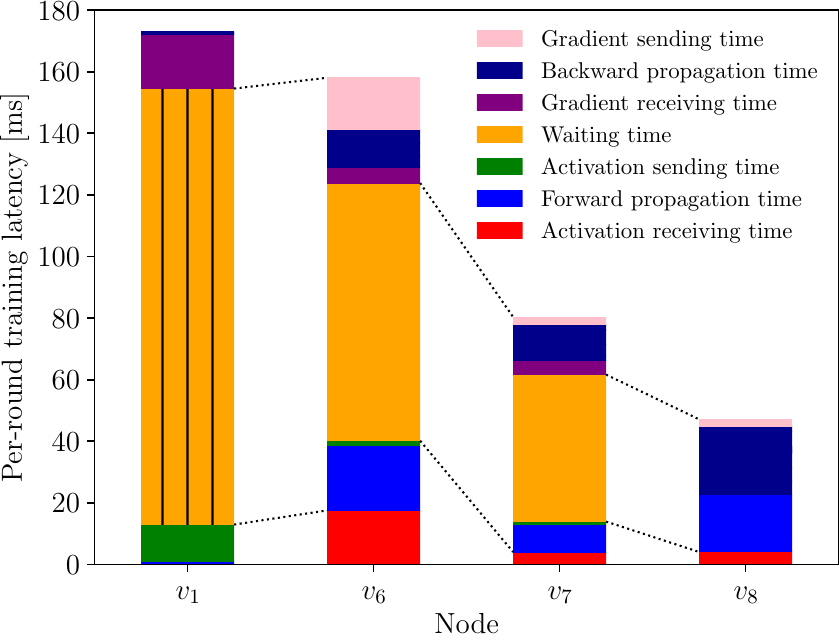}
    \subcaption{Proposed SFC-based MSL ($b=128$).}
    \label{fig:training_latency:proposal}
  \end{minipage}
  \begin{minipage}{0.49\hsize}
    \centering
    \includegraphics[width=\columnwidth]{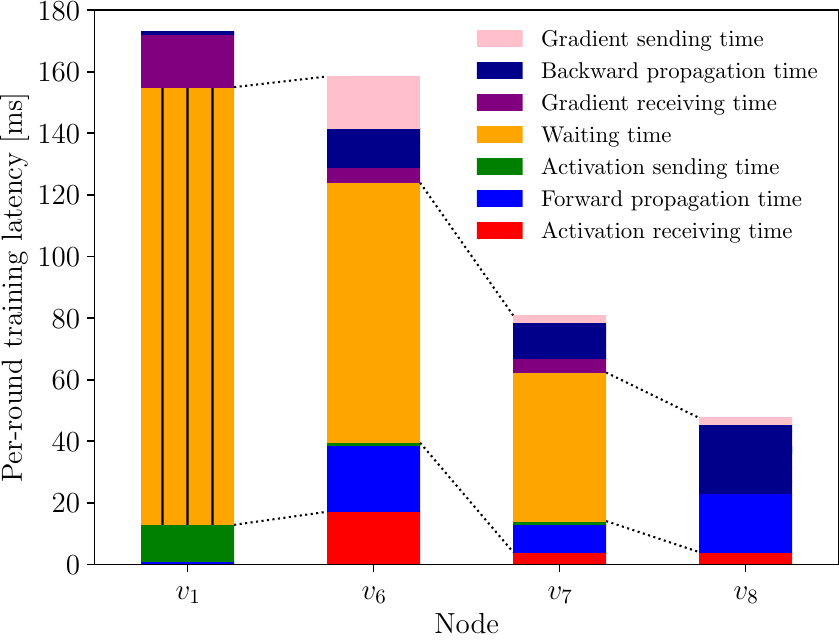}
    \subcaption{Baseline existing MSL ($b=128$).}
    \label{fig:training_latency:existing}
  \end{minipage}
  \caption{Comparison of per-round training latencies between the proposed SFC-based MSL and the baseline existing MSL.}
  \label{fig:training_latency}
\end{figure*}
\subsection{Training Latency}
This section evaluates the applicability of SFC to MSL by comparing the training latency of the proposed SFC-based MSL with that of the baseline existing MSL.
Figs.~\ref{fig:training_latency:proposal} and \ref{fig:training_latency:existing} illustrate the per-round training latencies of the proposed SFC-based MSL and the baseline existing MSL, respectively.
Similar to the inference latency, there is no significant difference in training latencies between the proposed SFC-based MSL and the baseline existing MSL.
Specifically, the per-round training latency for $v_1$ is measured at 173\,ms for both architectures when $b=128$.

We also confirm that the communication overhead depends on the batch size and the output dimension of each split sub-model, as shown in Table~\ref{table:spec_of_split_model}.
Compared to inference, training introduces additional challenges due to the presence of waiting time.
Specifically, during training, each sub-model must wait for the gradients from the subsequent sub-model after completing forward propagation.
This waiting time significantly increases the per-round training latency, as all nodes except the target server remain idle during this period.
The computational resources of these intermediate nodes are also underutilized.
One possible solution to address these issues is
adopting asynchronous training, where sub-models are updated less frequently and activations and gradients are exchanged only during selected epochs.

\subsection{Dynamic Path Reconfiguration Capability}
\label{sec:Dynamic Path Reconfiguration Capability}
\begin{figure}[!t]
  \centering
  \begin{minipage}{\hsize}
    \centering
    \includegraphics[width=\columnwidth]{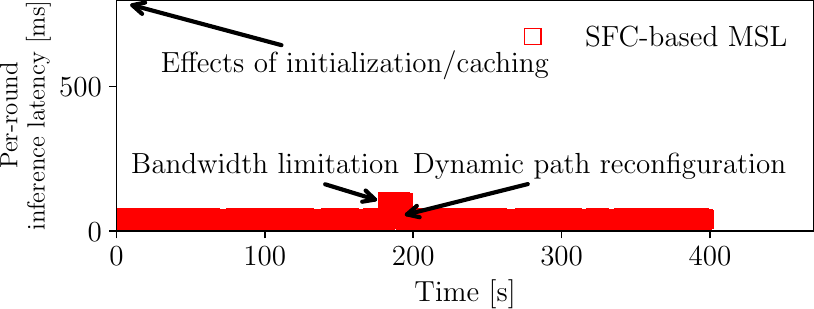}
    \subcaption{Proposed SFC-based MSI.}
    \label{fig:per-round_inference_latency_against_failure_event:proposal}
  \end{minipage}
  \begin{minipage}{\hsize}
    \centering
    \includegraphics[width=\columnwidth]{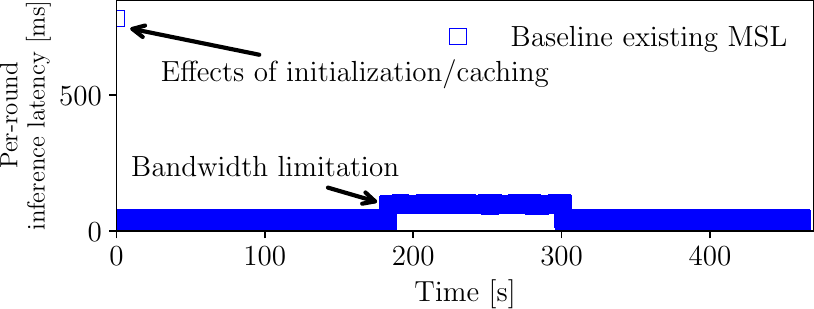}
    \subcaption{Baseline existing MSI.}
    \label{fig:per-round_inference_latency_against_failure_event:existing}
  \end{minipage}
  \caption{Variation in per-round inference latency over time ($b=1$).}
  \label{fig:per-round_inference_latency_against_failure_event}
\end{figure}
\begin{figure}[!t]
  \centering
  \begin{minipage}{\hsize}
    \centering
    \includegraphics[width=\columnwidth]{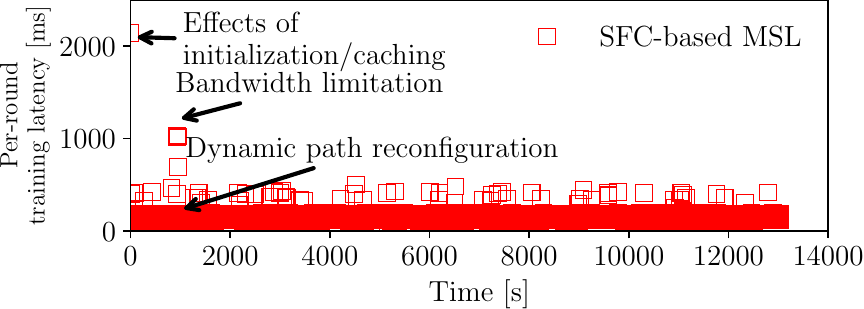}
    \subcaption{Proposed SFC-based MSL.}
    \label{fig:per-round_training_latency_against_failure_event:proposal}
  \end{minipage}
  \begin{minipage}{\hsize}
    \centering
    \includegraphics[width=\columnwidth]{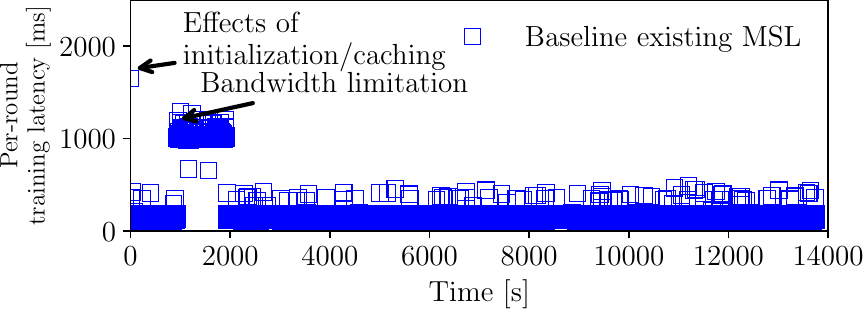}
    \subcaption{Baseline existing MSL.}
    \label{fig:per-round_training_latency_against_failure_event:existing}
  \end{minipage}
  \caption{Variation in per-round training latency over time ($b=128$).}
  \label{fig:per-round_training_latency_against_failure_event}
\end{figure}
This section evaluates the dynamic path reconfiguration capability of the proposed architecture.
In case of the proposed architecture, the SDN controller monitors the network condition of the link $(v_3, v_4)$ in Fig.~\ref{fig:evaluation_environment} at regular intervals of 1 second.
If the controller detects that the bandwidth of the link has remained below 1\,Mbps during inference or 10\,Mbps during training for the past 10 seconds, it identifies the link as congested and updates the segment list from $(v_6, v_7, v_5)$ to $(v_6, v_9, v_{10}, v_7, v_5)$.
This update creates a detour that bypasses the congested link $(v_3, v_4)$, forming a new service path: $v_1 \rightarrow v_2 \rightarrow v_3 \rightarrow v_6 \rightarrow v_3 \rightarrow v_9 \rightarrow v_{10} \rightarrow v_4 \rightarrow v_7 \rightarrow v_4 \rightarrow v_5 \rightarrow v_8$.
In contrast, The existing MSL is assumed to lack the capability to dynamically adjust the service path in response to network conditions.

To simulate network congestion, a bandwidth limitation of 1\,Mbps during inference or 10\,Mbps during training is applied to the link $(v_3, v_4)$ using the \texttt{tc} command after approximately 180 seconds for inference or 1,000 seconds for training.
This limitation is maintained for 100 seconds during inference or 1,000 seconds during training before restoring the original bandwidth of 1,000\,Mbps.

Figs.~\ref{fig:per-round_inference_latency_against_failure_event} and \ref{fig:per-round_training_latency_against_failure_event} show the variations in per-round inference latency and per-round training latency over time, respectively.
Initially, the per-round inference latency and training latency are approximately 800\,ms and 2,000\,ms, respectively, due to initialization and caching effects.
Both the per-round inference latency and training latency stabilize around 40\,ms and 170\,ms during the first 180 seconds and 1,000 seconds of processing, respectively.
When the bandwidth limitation is applied to the link $(v_3, v_4)$, both the per-round inference latency and training latency sharply increase to approximately 100\,ms and 1,000\,ms, respectively, reflecting the impact of congestion.
In the proposed SFC-based architecture, the SDN controller detects the congestion and dynamically updates the segment list to reroute traffic, enabling the system to recover its original performance level.
In contrast, the baseline existing architecture lacks this capability, resulting in persistently elevated latency until the bandwidth is restored.

\section{Conclusion}
\label{sec:Conclusion}
In this paper, we designed and implemented a Service Function Chaining (SFC) architecture tailored for Multi-hop Split Inference (MSI) and Split Learning (MSL), where split sub-models are treated as service functions and linked to form a service chain representing the global model.
As a proof of concept, we implemented Neural Service Functions (NSFs) as stateful network functions that execute split sub-models, leveraging Segment Routing over IPv6 (SRv6), transparent proxy (TPROXY), and extended Berkeley Packet Filter (eBPF)-based SFC proxy.
The evaluation results demonstrated that: (1) the proposed architecture is feasible for both MSI and MSL; (2) it is particularly suitable for real-time inference MSI scenarios with small mini-batch sizes; and (3) it supports dynamic path reconfiguration, enabling it to bypass congested links and maintain performance.

In future work, we plan to explore data compression techniques to reduce the size of smashed data, thereby mitigating communication overhead as the number of split points increases.
We also plan to develop asynchronous MSL over SFC to reduce the waiting time during training.
%
\IEEEpeerreviewmaketitle



\ifCLASSOPTIONcaptionsoff
  \newpage
\fi



\bibliographystyle{IEEEtran}
\bibliography{ref.bib}
\iftoc
\newpage
\tableofcontents
\newpage
\listoffigures
\listoftables
\listofalgorithms
\fi
\end{document}